\documentstyle[osa,aps,prb,epsfig]{revtex}
\begin{document}
\twocolumn[\hsize\textwidth\columnwidth\hsize\csname
@twocolumnfalse\endcsname
\title{Analyzing Intrinsic Superconducting Gap by Means of Measurement
of Bi$_2$Sr$_2$CaCu$_2$O$_{8+x}$ superconductors$^{\ast\ast\ast}$}
\author{Hyun-Tak Kim (htkim@etri.re.kr, ~kimht45@hotmail.com)}
\address{Telecom. Basic Research Lab., ETRI, Taejon 305-350, Korea}
\maketitle{}
\newpage
\begin{abstract}
For the mechanism of high-$T_c$ superconductivity in inhomogeneous
Bi$_2$Sr$_2$CaCu$_2$O$_{8+x}$ superconductors, we demonstrate the
intrinsic superconducting gap, $\triangle_i$, and pairing symmetry
by using a developed $\triangle=\triangle_i/\rho$, where
$\triangle$ is the observed energy gap and 0$<\rho\le$1 is band
filling. When $\rho$=1, $\triangle=\triangle_i\le10$ meV, measured
at a node, is intrinsic. When 0$<\rho<$1, $\triangle$ implies an
averaging of $\triangle_i$ over the measurement region, which is
an effect of the measurement. From spectra of the density of
states (DOS), $b=2\triangle_i/k_BT_c$ is less than 4 when $\rho$=1
and the DOS indicates $s$-wave symmetry. The superconducting gap
anisotropy is attributed to the inhomogeneity of the metal phase
and the insulating $d$-wave phase in the measurement region.\\ \\
\end{abstract}
]

To clarify the mechanism of high-$T_c$ superconductivity, it is
essential to know the intrinsic superconducting energy gap and the
intrinsic density of states (DOS). High-$T_c$ superconductors are
intrinsically spatially inhomogeneous; this is attributed to
metal-insulator instability in which a half-filled metal lies in
an unstable position at the local charge-density-wave
potential.$^{1)}$ The instability implies that the probability of
one electron occupying a cubic unit is less than one. This
indicates that a metallic system is separated into metal and
insulator phases composed of cubic units with and without an
electron, respectively, because the electron is not divided. Pan
$et ~al.^{2)}$ and Wang $et ~al.^{3)}$ have suggested as an
alternative explanation that nonlinear screening of the ionic
potential leads to strong inhomogeneous redistribution of the
local hole density. The inhomogeneity was proven experimentally by
scanning tunneling microscopy (STM).$^{2),4),5),6)}$ Both the
local DOS and the energy gap are correlated spatially and vary on
the surprisingly short length scale of about $14~\AA$.$^{2),3)}$
The transitions from superconducting state to a low-temperature
pseudogap state are spatially continuous and occur on a very local
scale of 1-3 nm. The smallest characteristic size of
superconducting islands is about 3 nm in diameter.$^{4),5),6)}$

In the inhomogeneous superconductors, which are a mixture of an
insulating phase and a metal phase (superconducting phase at low
temperatures), the metallic (or insulating) characteristics are
measured by the average of the two phases, (Fig. 1). The measured
data are always composed of the effects of two phases,
consequently, the intrinsic characteristics in the respective
phases cannot be measured exactly. Therefore, since the discovery
of high-$T_c$ superconductors, problems such as the pseudogap, the
intrinsic superconducting energy gap, the intrinsic DOS, and
pairing symmetry remain unsolved although numerous experimental
data and theories have been obtained and developed.

In this paper, we analyze the intrinsic superconducting gap and
pairing symmetry by evaluating the local carrier density (or band
filling), with spectra measured by STM and angle-resolved
photoemission spectroscopy (ARPES) for
Bi$_2$Sr$_2$CaCu$_2$O$_{8+x}$(Bi-2212) superconductors. STM and
ARPES are very powerful tools for determining the most homogeneous
spatial regions in a crystal and for observing the intrinsic
effect, respectively.

\begin{figure}
\centerline{\epsfysize=9.0cm\epsfxsize=8.0cm\epsfbox{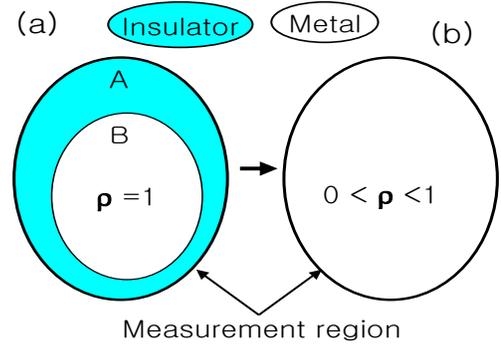}}
\vspace{-2.0cm} \caption{(a) An inhomogeneous superconductor with
both an insulator phase of the hump structure regarded as
pseudogap and a metal phase (superconducting phase at low
temperatures). (b) When the measurement region is observed, the
metal phase in region B is changed into an averaged metal phase
with 0$<\rho<$1 in Fig. (b). When only region B is measured, the
intrinsic gap and the intrinsic $T_c$ are obtained.}
\end{figure}

The well-known BCS DOS of ${\it Eq}$. (2), which is applied to a
homogeneous superconductor and defined at 0~K, cannot be applied
to experimental data measured in an inhomogeneous superconductor
(Fig. 1), which does not define k-space; therefore the DOS must be
converted into that of a homogeneous superconductor. Here, the
metal phase on homogeneous superconductors implies that there is
no charge difference between nearest neighbor sites, for example,
when there is one electron per atom in the electronic structure.
When the carriers in the metal phase in the inhomogeneous
superconductor are averaged over all atomic (or lattices) sites in
the measurement region, it is possible to change the inhomogeneous
superconductor into a homogeneous one with a carrier of an
effective charge.$^{7),8)}$ The effective charge of the carrier is
given as a fractional charge, $e'={\rho}e$, where
0$<{\rho}(=n/l){\le}$1 is band filling, $n$ is the number of
carriers in the metal phase (region B in Fig. 1) and $l$ is the
number of lattices.$^{7),8)}$ The number of bound charges,
$n_b=l-n$, is bound in the insulating phase with a pseudogap
(region A in Fig. 1); the total charges are conserved even in the
inhomogeneous superconductor. The effective charge is justified
only by means of measurement, that is, when not measured, the
effective charge becomes the elementary true charge in the metal
phase.

In the tunneling conductance ($\frac{dI}{dV}$), the observed
energy gap, $\triangle$, is given by

\begin{eqnarray}
{\triangle}=eV_{bias}={\triangle}_i/{\rho} ,
\end{eqnarray}

by substituting $e$ with $e'$. ${\triangle}$ increases as $\rho$
decreases and is an average value of the intrinsic energy gap,
${\triangle}_i$, over a measured region. ${\triangle}_i$ is
attributed to pairing of two electrons of true charge when
$\rho$=1, is constant irrespective of the extent of $\rho$, and is
determined by the minimum bias voltage. Moreover, we apply ${\it
Eq}$. (1) to the tunneling spectra and the photoemission spectra.

\begin{figure}
\vspace{-4.0cm}
\centerline{\epsfysize=10.0cm\epsfxsize=8.0cm\epsfbox{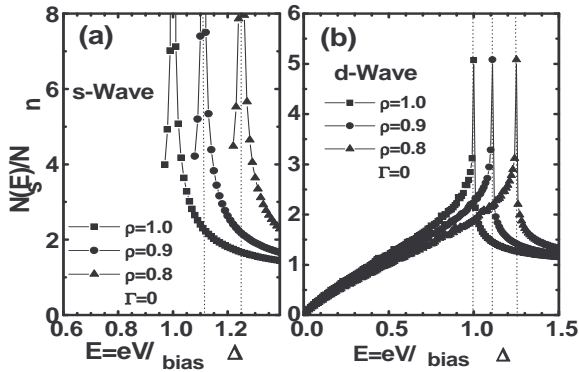}}
\vspace{0.5cm} \caption{(a) Band filling, $\rho$, dependence of
the $s$-wave DOS of $Eq$. (3) $\frac{N_s(E)}{N_n}=0$ is assumed in
$\vert\frac{(E=eV_{bias})}{\triangle}\vert<1$. Peak curves when
$\vert\frac{eV_{bias}}{\triangle}\vert<1$ are given for a
comparison to curves when
$\vert\frac{eV_{bias}}{\triangle}\vert>1$. (b) Band filling
dependence of the $d$-wave DOS of $Eq$. (3). The DOS were
precisely calculated from
$\frac{N_s(E)}{N_n}=\frac{1}{2\pi}{\int}{\frac{N_s(E,\phi)}{N_n}}d{\phi}$
 with ${\triangle}={\triangle}_scos(2{\phi})$ and $\Gamma=0$
by numerical analysis not using complete elliptic integrals.
Divergences in calculations were ignored because the divergences
have no derivative width (${d\phi}=0$).}
\end{figure}

The BCS DOS, tunneling conductance in the inhomogeneous
superconductor, is given by

\begin{eqnarray}
N_s(E)/N_n &=&Re\left(\frac{E}{\sqrt{E^2 -
{\triangle}^2}}\right),\\
 &=&Re\left(\frac{E}{\sqrt{E^2 - ({\triangle}_{i}/\rho)^2}}\right),
\end{eqnarray}

where ${\triangle}_i$ is ${\triangle}_s$ in a $s$-wave
superconductor, ${\triangle}_s$cos(2$\phi$) in a $d$-wave
superconductor, and $E$ is an applied bias voltage. Considering
the broadening effect, $E$ is substituted with $E=E-i\Gamma$ where
$\Gamma$ is the broadening parameter.$^{9)}$ Equation (3) is valid
in ${\vert}E{\vert}\ge\triangle$ in the $s$-symmetry case and is
averaged by phase angle $\phi$ in the $d$-symmetry case. Equation
(3) has a coherence peak at the energy gap which increases with
decreasing $\rho$, (Fig. 2); thus, the unsolved
problem$^{10)-12)}$ in the tunneling conductance, is explained. It
was experimentally demonstrated that analogous to Bi-2212, the
energy gap in Bi-2201 monotonically increases with a decreasing
hole concentration.$^{13)}$ Equation (3) and the coherence peak
energy when $\rho$=1 are regarded as the intrinsic DOS and
${\triangle}_i$, respectively. Equation (3) and the peak energy
when $0<\rho<1$ imply an averaging of the intrinsic DOS and
${\triangle}_i$ over the measurement region, which is the effect
of the measurement.

Pan $et ~al.^{2)}$ measured the well-shaped tunneling conductance
curve 5 with a clear coherence peak at ${\triangle\approx}$25 meV,
using STM, as shown in Fig. 3 (c) of their paper. On the basis of
their analysis, curve 5, which is observed at a larger $\rho$
value than other curve numbers, resembled the spectra behavior of
an oxygen over-doped Bi-2212 crystal. This implies that the
intrinsic DOS curve is similar to the curve measured in over-doped
crystals. Similar analyses were presented by several other groups
as well.$^{4)-6)}$ Hasegawa $et ~al.^{14)}$ and Kitazawa $et
~al.^{15)}$ observed tunneling conductances for single crystals of
Bi-2212 with ${\it atomic ~resolution}$ which can be regarded as
$\rho\approx$1, using STM. The conductances revealed clear
coherence peaks at about ${\triangle\approx}$22 meV$^{14)}$ and at
about ${\triangle\approx}$17 meV as shown in Fig. 6 of Kitazawa
$et ~al.$'s paper$^{15)}$, with flat bottom regions around
$V_{bias}$=0. The broadening parameter $\Gamma$ is less than
1$\%$. They suggested that the observation is favored by the
$s$-wave pairing mechanism. Moreover, analysis results for YBCO
were the same as those for Bi-2212.$^{14)}$ Using intrinsic
tunneling spectroscopy for high-quality Bi-2212 crystals and
films, the energy gaps were measured as $2{\triangle}{\approx}25$
meV by Lee and Iguchi$^{16)}$ and as an inter-branch value,
2${\triangle}{\approx}$25 meV, from I-V curves by Doh $et
~al.^{17)}$.

On the other hand, the photoemission spectra imply the same DOS as
the tunneling conductance. The measured spectra have a much larger
insulating effect than spectra measured by STM because the size of
the X-ray beam cannot be reduced to less than 30~$\AA$. However,
angle-resolved PES gives information on DOS at a node and a
non-node.

Although the gap anisotropy has been suggested as evidence of
$d$-wave symmetry,$^{18)}$ on the contrary it indicates, according
to ${\it Eq}$. (1), that $\rho_{\Gamma-Y}$ for the small gap at a
node ${\Gamma-Y}$ is larger than $\rho_{\bar{M}}$ for the large
gap at $\bar{M}$ of ($\pi$,0). The large
$\rho_{\Gamma-Y}(=\rho_{\bar{M}}+\frac{n_b}{l}\approx1)$ is
because bound charges, $n_b$, become carriers when the hump
structure in the spectra disappears at the node $\Gamma-Y$;
transition from insulator (region A in Fig. 1) with the pseudogap
to metal occurs. Figure 1 of Shen $et ~al.$'s paper$^{18)}$ showed
the absence of the hump at the node, which indicates that the
pseudogap has $d$-wave symmetry. The gap anisotropy is the effect
of measurement and the gap anisotropy of $\triangle_i$ cannot be
measured for inhomogeneous superconductors. A clear coherence peak
at the node with $\rho\approx$1 was observed with a small gap,
${\triangle\le}$10 meV,$^{19)-21)}$ and the small gap is much
closer to $\triangle_i$ than the large gap at $\bar{M}$;
${\triangle_i\le} 10$ meV. Note that the energy gap observed by
STM is larger than the small gap observed at the node by PES,
because the averaged STM gap over the Fermi surface is an average
of the small gap at a node and the large gap at a non-node. The
observed dip-hump structure in the tunneling spectra$^{13)}$ comes
from the non-node $\bar{M}$. Furthermore, when the hump structure
disappears at nodes and non-nodes, the gaps become isotropic due
to the same $\rho$. The isotropic gaps in the structures without
humps were experimentally observed, although the observation
depended on the properties of the cleaned sample surface$^{22)}$;
the metal phase with $\rho$ = 1 at the sample surface is unstable
because of the metal-insulator instability$^{1)}$. Electronic
Raman scattering also showed the isotropic gaps measured in an
overdoped Bi-2212 crystal.$^{23)}$ The isotropic gap is evidence
of $s$-wave symmetry.

The strong spin-fermion model$^{24)}$ for a homogeneous $d$-wave
superconductor of $\rho = 1$ proposed both the dip-hump structure
and the broad coherence peak. However, according to the model,
although the coherence peak and the dip-hump structure should
disappear at the same time at the node $\Gamma-Y$, only the
dip-hump structure disappeared$^{18)}$, while the coherence peak
remained$^{19)-21)}$. This indicates that the model does not
explain the experimental data.

The paramagnetic Meissner effect of the Josephson-$\pi$
junction,$^{25)}$ suggested as evidence of $d$-wave symmetry, was
revealed to be the effect of a trapped flux.$^{26)}$

Wollman and co-workers$^{27),28)}$ observed the integer flux
(1$\Phi_0$) and a Fraunhofer modulation pattern as evidence of
$s$-wave symmetry and the half flux ($\frac{1}{2}\Phi_0$)and a dip
modulation pattern at zero field as evidence of the
Josephson-$\pi$ junction of $d$-wave symmetry, simultaneously,
using a corner SQUID for YBCO single crystals. Current in the
corner SQUID passes through a (110) plane showing $d$-wave
characteristics while cornering. Evidence of $d$-wave symmetry
came from the $d$-wave insulating phase (region A in Fig. 1) when
compared with the result of the angle-resolved PES as mentioned
previously, although Wollman and co-workers indicated that
experiments were performed with crystals with a single phase of
$\rho$=1.

The zero-bias conductance peak (ZBCP) as evidence of $d$-wave
symmetry, theoretically suggested by Kashiwaya et al.$^{29)}$ and
Tanaka and Kashiwaya$^{30)}$, was observed in a thin film and an
under-doped crystal by tunneling experiments.$^{31)-33)}$ They
assumed that the crystals used had a single phase with $\rho$=1.
However, this author insists that the ZBCP came from the $d$-wave
insulating phase for the inhomogeneous superconductors, because
Kohen $et~al.^{34)}$ could not observe the ZBCP on overdoped [110]
oriented Y$_{0.8}$Ca$_{0.2}$Ba$_2$Cu$_3$O$_{7-\delta}$
films$^{34)}$ by planar tunneling or point contact measurements;
overdoped crystals have only slight or no $d$-wave insulating
phase$^{35)}$. In addition, Ekino $et~al.^{36)}$ indicated that
the ZBCP observed with break-junction tunneling is attributed to
the interface of the break junction.

In conclusion, only in the metal phase (superconducting phase at
low temperatures) of region B in Fig. 1 (a),
$b=2{\triangle}_i/k_BT_{c,max}$ is applied, where $T_{c,max}$ is
the intrinsic critical temperature of the maximum measured value.
Here, note that $T_c$ in region B has the maximum value
irrespective of the extent of the metal phase with $\rho$=1
because of the largest DOS$^{7),8)}$. When $T_{c,max}{\approx}$93
K observed at an optimal doping and ${\triangle_i}\approx$10 meV
observed at the node are used, the coupling constant, $b$, is less
than 4.0. The intrinsic DOS showed $s$-wave symmetry. Thus,
high-$T_c$ superconductivity can be explained within the context
of the BCS theory.

We acknowledge W. Sacks, S. H. Pan, K. Lee, and Y.-J. Doh for
communications. We thank N. Miyakawa, W.-J. Kim, T. Nishio, and
K.-Y. Kang for their valuable comments.


\begin{references}
\bibitem[\ast\ast\ast]{} J. Phys. Soc. Jpn. Vol. {\bf 71} No. 9, (2002) 2106.
\bibitem[1)]{} Hyun-Tak Kim : Phys. Rev. B {\bf 54}, (1996) 90.
\bibitem[2)]{} S. H. Pan $et ~al.$ : Nature {\bf 413},(2001) 282.
\bibitem[3)]{} Z. Wang, J. R. Engelbrecht, S. Wang, H. Ding, and S. H.
Pan : Phys. Rev. B {\bf 65}, (2002) 64509.
\bibitem[4)]{} T. Cren, D. Roditchev, W. Sacks, and J. Klein :
Eryophys. Lett. {\bf 54(1)}, (2001) 84.
\bibitem[5)]{} C. Howald, P. Fournier, and A. Kapitulnik : Phys. Rev. B {\bf 64},(2001) 100504(R).
\bibitem[6)]{}K. M. Lang, V. Madhavan, J. E. Hoffman, E. W. Hudson, H.
Eisaki, S. Uchida, and J. C. Davis : Nature {\bf 415}, (2002) 412.
\bibitem[7)]{} Hyun-Tak Kim : Physica C {\bf 341-348}, (2000) 259.
\bibitem[8)]{} Hyun-Tak Kim : New Trends in Superconductivity
(Ed. J.F. Annett and S. Kruchinin, Kluwer, 2002), NATO Science
Series Vol. 67, p. 137; cond-mat/0110112.
\bibitem[9)]{} R. C. Dynes, V. Narayanamurti, and J. P. Garno : Phys.
Rev. Lett. {\bf 41}, (1978) 1509.
\bibitem[10)]{} N. Miyakawa, J. F. Zasadzinski, L. Ozyuzer, P. Guptasarma, D. G. Hinks, C. Kendziora,
and K. E. Gray : Phys. Rev. Lett. {\bf 83}, (1999) 1018.
\bibitem[11)]{} N. Miyakawa, P. Guptasarma , J. F. Zasadzinski, D. G. Hinks, and K.
E. Gray : Phys. Rev. Lett. {\bf 80}, (1998) 157.
\bibitem[12)]{} C. Renner, B. Revaz, J.-Y. Genoud, K. Kadowaki, and O.
Fisher : Phys. Rev. Lett. {\bf 80}, (1998) 149.
\bibitem[13)]{} N. Miyakawa, J. F. Zasadzinski, M. Asano, D. Henmi,
S. Oonuki, K. Sasaki, T. Kaneko, L. Ozyuzer, and K. E. Gray :
Physica C {\bf 357-360}, (2001) 126.
\bibitem[14)]{} T. Hasegawa, M. Nantoh, A. Takagi, W. Yamaguchi, M. Ogino,
M. Kawasaki, J. P. Gong, H. Koinuma, and K. Kitazawa : Physica B
{\bf 197}, (1994) 617.
\bibitem[15)]{} K. Kitazawa, T. Hasegawa, and H. Sugawara : J. Korean
Phys. Soc. {\bf 31}, (1997) 27.
\bibitem[16)]{} K. Lee and I. Iguchi : Physica C {\bf 367}, (2002) 376.
\bibitem[17)]{} Y. -J. Doh, H. -J. Lee, and H. -S. Chang : Phys. Rev.
B {\bf 61}, (2000) 3620.
\bibitem[18)]{} Z.-X. Shen $et ~al.$ : Phys. Rev. Lett. {\bf 70}, (1993) 1553.
\bibitem[19)]{} A. Kaminski $et ~al.$ : Phys. Rev. Lett. {\bf 21}, (2000)
1788.
\bibitem[20)]{} T. Valla $et ~al.$ : Science {\bf 285}, (1999) 2110.
\bibitem[21)]{} R. J. Kelley, C. Quitmann, M. Onellion, H. Berger, P.
Almeras, and G. Margaritondo : Science {\bf 271}, (1996) 1255.
\bibitem[22)]{} H. Ding, J. C. Campuzano, K. Gofron, C. Gu, R. Liu,
B. W. Veal, and G. Jennings : Phys. Rev. B {\bf 50}, (1994) 1333.
\bibitem[23)]{} C. Kendziora, R. J. Kelley, and M. Onellion : Phys. Rev.
Lett. {\bf 77}, (1996) 727.
\bibitem[24)]{} A. Abanov and A. V. Chubukov : Phys. Rev. B {\bf 61}, (2000) R9241.
\bibitem[25)]{} M. Sigrist and T. M. Rice : J. Phys. Soc. Jpn. {\bf 61},
 (1992) 4283.
\bibitem[26)]{} Hyun-Tak Kim, H. Minami, W. Schmidbauer, J. W. Hodby,
A. Iyo, F. Iga, and H. Uwe : J. Low Tem. Phys. {\bf 105(3/4)},
(1996) 557; cond-mat/0206432.
\bibitem[27)]{} D. A. Wollman, D. J. Van Harlingen, W. C. Lee, D. M.
Ginsberg, and A. J. Leggett : Phys. Rev. Lett. {\bf 71}, (1993)
2134.
\bibitem[28)]{} D. A. Wollman, D. J. Van Harlingen, J. Giapintzakis, and D.
M. Ginsberg : Phys. Rev. Lett. {\bf 74}, (1995) 797.
\bibitem[29)]{} S. Kashiwaya, Y. Tanaka, M. Koyanagi, H. Takashima,
and K. Kajimura : Phys. Rev. B {\bf 51}, (1995) 1350.
\bibitem[30)]{} Y. Tanaka and S. Kashiwaya : Phys. Rev. Lett. {\bf 74},
(1995) 3451.
\bibitem[31)]{} I. Iguchi, W. Wang, M. Yamazaki, Y. Tanaka, and S.
Kashiwaya : Phys. Rev. B {\bf 62}, (2000) R6131.
\bibitem[32)]{} J.Y.T. Wei, N.-C Yeh, W.D. Si, and X.X. Xi : Physica B {\bf 284-288}, (2000) 973.
\bibitem[33)]{} J.Y.T. Wei, N.-C. Yeh, D.F. Garrigus, and M.
Strasik : Phys. Rev. Lett. {\bf 81}, (1998) 2542.
\bibitem[34)]{} A. Kohen, Y. Dagan, and G. Deutscher : Physica C
{\bf 341-348}, (2000) 687.
\bibitem[35)]{} J. L. Tallon and J. W. Loram : Physica C {\bf 349},
(2001) 53.
\bibitem[36)]{} T. Ekino, Y. Sezaki, and H. Fujii : Phys. Rev. B {\bf
60}, (1999) 6916.

\end{references}
\end{document}